\documentclass[aip, jcp, amsmath, amsthm, amssymb, citeautoscript, reprint]{revtex4-2}
\usepackage{graphicx, float, xcolor, pgf}
\usepackage{physics, siunitx}
\usepackage[caption=false, justification=centerlast]{subfig}
\usepackage{gensymb}
\usepackage{hyperref}
\hypersetup{
    pdfauthor={Devansh Sharma and Amartya Bose},
    pdftitle={Impact of Loss Mechanisms on Linear Spectra of Excitonic and Polaritonic Aggregates},
    colorlinks=true,
}

\allowdisplaybreaks

\begin{document}
\title{Impact of Loss Mechanisms on Linear Spectra of Excitonic and Polaritonic Aggregates}
\author{Devansh Sharma}
\affiliation{Department of Chemical Sciences, Tata Institute of Fundamental Research, Mumbai 400005, India}
\author{Amartya Bose}
\email{amartya.bose@tifr.res.in}
\affiliation{Department of Chemical Sciences, Tata Institute of Fundamental Research, Mumbai 400005, India}
\begin{abstract}
    The presence of loss mechanisms governed by empirical time-scales affect the
    dynamics and spectra of systems in profound ways. However, incorporation of
    these effects and their interaction with the thermal dissipative
    environments interacting with the system prove to be challenging. We have
    recently developed the path integral Lindblad dynamics (PILD) method to
    combine numerically rigorous path integral simulations with Lindblad
    dynamics to account for such empirical loss mechanisms. In this work, we
    utilize the PILD method to study the absorption and circular dichroism
    spectra of chiral molecular aggregates and excitonic polaritons. We
    demonstrate that the effect of loss on particular states in both systems can
    differ not just on the basis of the symmetries of the state but also on the
    basis of complicated ``interactions'' of the system and the loss mechanism
    with the dissipative environments. We present probably the first numerical
    exploration of the CD spectrum of chiral molecular aggregates confined in a
    cavity. While the CD spectrum of just the excitonic aggregates itself is not
    amenable to simplistic understanding like the exciton chirality (EC) rule,
    the CD spectrum of polaritonic molecules is even more complex. Additionally,
    the impact of empirical loss on the polaritonic CD spectrum seems to be
    highly site-dependent. The impact of a lossy cavity is qualitatively
    different from the impact of a molecule that leaks the excitation. We
    explore some of those effects in depth leveraging the framework of path
    integral Lindblad dynamics.
    
\end{abstract}
\maketitle
\section{Introduction}
Spectroscopy is one of the most basic tools for characterizing chemical systems.
While absorption spectra gives an insight into the electronic structure, circular
dichroism spectra provides insights into the geometry of chiral systems. These
spectra are broadened due to the coupling of the eigenstates states to
environment degrees of freedom which are usually thermally populated.
Simulations of the exact lineshape proves to be challenging owing to the
continuous manifold of environmental states participating in the dynamics. The
presence of loss mechanisms in many systems additionally broadens the spectral
features, which poses an additional challenge to accurate predictions of these
broadened line-shapes. Methods based on wave functions such as density matrix
renormalization group~\cite{whiteDensityMatrixFormulation1992,
schollwockDensitymatrixRenormalizationGroup2005,
schollwockDensitymatrixRenormalizationGroup2011,
paeckelTimeevolutionMethodsMatrixproduct2019} or multiconfiguration
time-dependent Hartree~\cite{meyerMulticonfigurationalTimedependentHartree1990}
and its multilayer
formulation~\cite{wangMultilayerFormulationMulticonfiguration2003,
wangMultilayerMulticonfigurationTimeDependent2015} have been pushing the limits
of what is computationally tractable, however, they still are inadequate in
describing the dynamics in the condensed phase where a continuum of environment
degrees of freedom are thermally populated.

Methods that simulate the reduced density matrix (RDM) of the system overcome
many of these challenges by integrating out an appropriately selected set of
environment modes. This tracing over the environment makes the dynamics
non-Markovian. Various approximate methods like the Bloch-Redfield master
equation~\cite{blochGeneralizedTheoryRelaxation1957,
    redfieldTheoryRelaxationProcesses1957} and the Lindblad or the
Gorini-Kossakowski-Sudarshan-Lindblad master
equation~\cite{lindbladGeneratorsQuantumDynamical1976,
    goriniCompletelyPositiveDynamical1976} treat the bath perturbatively or make
a Markovian approximation for the dynamics. Mixed quantum classical methods like
surface hopping~\cite{tullyMolecularDynamicsElectronic1990,
    tullyTrajectorySurfaceHopping2003} treat the environment classically. However, the
results obtained from these methods cannot be improved systematically.
Feynman's path integral approach~\cite{feynmanQuantumMechanicsPath2010} provides
an alternate route to simulating such systems that is numerically rigorous. The
enviornment modes are incorporated using the Feynman-Vernon influence
functional~\cite{feynmanTheoryGeneralQuantum1963}. The hierarchical equation of
motion~\cite{tanimuraQuantumClassicalFokkerPlanck1991,
    tanimuraNumericallyExactApproach2020} (HEOM) and the quasi-adiabatic propagator
path integral~\cite{makriTensorPropagatorIterativeI1995,
    makriTensorPropagatorIterativeII1995} (QuAPI) and later developments on
them~\cite{cerrilloNonMarkovianDynamicalMaps2014,
    rahmanChebyshevHierarchicalEquations2019, xuTamingQuantumNoise2022,
    strathearnEfficientNonMarkovianQuantum2018, jorgensenExploitingCausalTensor2019,
    makriSmallMatrixDisentanglement2020, boseMultisiteDecompositionTensor2022,
    boseImpactSolventStatetoState2023, boseQuantumCorrelationFunctions2023,
    yanEfficientPropagationHierarchical2021} provide lucrative ways of implementing
path integral for large open quantum systems. Recent applications on large
biomolecular aggregates~\cite{yanEfficientPropagationHierarchical2021,
    kunduB800toB850RelaxationExcitation2022, boseTensorNetworkPath2022,
    boseImpactSpatialInhomogeneity2023} show the full power and applicability of
these methods in simulating large systems.

However, it is generally difficult to incorporate loss or gain processes
described by empirical time scales into numerically exact simulations. These
empirical loss mechanisms are, in fact, quite ubiquitous. Consider, for example,
the loss of an exciton at particular sink sites to the ``special pair'' in
photosynthetic aggregates or the loss of photons from leaky cavities. It is not
always possible to characterize such processes using proper spectral densities
and harmonic baths. Use of a fictitious bath can often induce artifacts stemming
from particular functional form used. While non-Hermitian descriptions can be
used in some cases, the resultant dynamics is non-unitary leading to spurious
effects in correlation functions and spectra. The Lindblad master equation,
which maintains the unitarity of evolution, offers a lucrative alternative to
these non-Hermitian descriptions in simulating such systems. Now, while the
Lindblad approach is good for the lossy modes, we do not want to treat the
effect of the thermal vibrations using Lindblad dynamics. A combination of the
fully semiclassical partial linearized density
matrix~\cite{huoCommunicationPartialLinearized2011,
provazzaSemiclassicalPathIntegral2018} with Lindblad master equation has been
used to study the effect of losses on the 2D electronic spectra of polaritonic
aggregates by Mondal \textit{et al}~\cite{mondalQuantumDynamicsSimulations2023}.
Recently, one of us has developed an efficient approach towards combining
numerically exact QuAPI simulations, accounting for the bath effects, with the
Lindblad master equation to account for such empirical loss mechanisms at no
extra cost~\cite{boseIncorporationEmpiricalGain2024}. This path integral
Lindblad dynamics (PILD) approach guarantees a positive definite dynamical map
for the dynamics. Being based on QuAPI, it is able to handle arbitrary thermal
environments with equal ease. In addition different loss mechanisms can be
studied in parallel without having to rerun the path integral simulation. The
impact of the loss of an exciton on the dynamics in the Fenna--Matthew--Olson
complex was explored~\cite{boseIncorporationEmpiricalGain2024}.

Here we extend the PILD approach to simulate the line-shapes of linear spectra
of aggregates in presence of loss and attempt to answer the natural question:
how does the spectrum of a system get affected by the presence of these loss
mechanisms? These losses become especially interesting in cases of excitonic
molecular wires where the exciton is lost at one site, and excitonic polaritons
where the cavity is known to be lossy. In Sec.~\ref{sec:method}, we summarize
the path integral Lindblad dynamics
framework~\cite{boseIncorporationEmpiricalGain2024} and extend it to allow the
study of linear spectra. In Sec.~\ref{sec:results}, we describe the chiral
molecular aggregates that are of interest. We explore the absorption and CD
spectra under different loss cases. While one might be tempted to make simple
generalizations about the impact of loss on the absorption and CD spectra, it
seems that little can be simply stated beyond the fact that loss, in general,
broadens peaks and decreases their intensities. We show that which peak is
broadened and by what amount is a non-trivial problem. We come up with rules of
thumb, which while still not complete, gives a better zeroth order approximate
understanding of the phenomenon. We demonstrate the caveats involved by probing
deeper. PILD allows for a simple incorporation of processes described by
empirical timescales in rigorous numerical simulations for the first time.
Finally, we end by providing some concluding remarks and future directions in
Sec.~\ref{sec:conclusions}.

\section{Method}\label{sec:method}
Consider an open quantum system described by a system-environment decomposed
Hamiltonian,
\begin{align}
    H & = H_0 + H_\text{env},\label{eq:Htot}
\end{align}
where $H_0$ is the system Hamiltonian and $H_\text{env}$ is the Hamiltonian
describing the environment and its interaction with the system. For a typical
exciton transport system, the system can be defined in terms of the singly
excited subspace spanned by $\ket{j}$ representing the excitation occurring on the
$j$th molecule with all other molecules in the ground state. The resultant
Hamiltonian has the well-known Frenkel form:
\begin{align}
    H_0 &= \sum_j \epsilon_j \dyad{j} + \sum_{jk} h_{jk}\left(\dyad{j}{k} + \dyad{k}{j}\right),\label{eq:H0_ex}
\end{align}
where $\epsilon_j$ is the energy of the locally excited state on the $j$th
molecule and $h_{jk}$ is the amplitude for the excitation to hop from the $j$th
to the $k$th molecule. Notice that this Hamiltonian has been expressed in the
so-called first-excitation subspace. For the simulation of spectra, it will be
necessary to augment the basis with the electronic ground state, $\ket{0}$,
which has all the monomers in the ground state.

The environment includes the thermal solvent to which the system is exposed and
the molecular vibrations of the system. Under linear response, these modulations
can be mapped on to baths of harmonic oscillators that locally interact with a
monomer,
\begin{align}
    H_\text{env} &= \sum_j \sum_b \frac{p_{jb}^2}{2} + \frac{1}{2}\omega_{jb}^2 x_{jb}^2 - c_{jb}x_{jb}\dyad{j},
\end{align}
where $\omega_{jb}$ is the frequency of the $b$th harmonic mode that interacts
with the $j$th molecule and $c_{jb}$ is the corresponding coupling. The mapping
of a molecular solvent onto this set of harmonic baths is done through the bath
spectral density
\begin{align}
    J_j(\omega) & = \frac{\pi}{2}\sum_b \frac{c_{jb}^2}{\omega_{jb}}\delta(\omega-\omega_{jb}).\label{eq:spectral_density}
\end{align}
For a molecular solvent, $J_j(\omega)$ can be calculated using molecular dynamics
simulations~\cite{olbrichTheorySimulationEnvironmental2011,
maityDFTBMMMolecular2020}. In addition to this environment, the system also has
loss mechanisms which cannot be described in this Hamiltonian formalism. These
would be incorporated using the Lindblad master equation~\cite{lindbladGeneratorsQuantumDynamical1976,goriniCompletelyPositiveDynamical1976}. Therefore, we need a
way to simulate the reduced density matrix of the system in the presence of both
the thermal environment and the Lindbladian jump operators $L_j$.

According to the path integral Lindblad
dynamics~\cite{boseIncorporationEmpiricalGain2024} approach, the loss mechanisms
are incorporated using a modified Nakajima-Zwanzig master
equation~\cite{nakajimaQuantumTheoryTransport1958,
zwanzigEnsembleMethodTheory1960}. If the initial state is uncorrelated, $\rho(0)
= \tilde\rho(0)\otimes\exp(-\beta H_\text{env})/Z$, then the time-evolution of
the reduced density matrix corresponding to the system is given by:
\begin{align}
    \dot{\Tilde{\rho}}^{(L)}(t) & = -\frac{i}{\hbar}\mathcal{L}_0 \Tilde{\rho}^{(L)}(t) + \int_{0}^{\tau_{\mathrm{mem}}}\mathcal{K}(\tau)\Tilde{\rho}^{(L)}(t-\tau)d\tau \nonumber \\
    &+ \sum_j \left(L_j\,\Tilde{\rho}^{(L)}(t)\,L_j^\dag - \frac{1}{2} \acomm{L_j^\dag L_j}{\Tilde{\rho}^{(L)}(t)}\right)\label{eq:piLD_diff}
\end{align}
where $\Tilde{\rho}^{(L)}(t)$ is the time-evolved reduced density matrix for the open
quantum system under the influence of the Lindbladian jump operators,
$\mathcal{L}_0=\comm{H_0}{\,\cdot\,}$ is the system Liouvillian, $\mathcal{K}$ is
the memory kernel corresponding to the thermal environment, and
$\tau_{\mathrm{mem}}$ is the memory length.

Obtaining the memory kernel, $\mathcal{K}$, is extremely challenging. While
various ways of approximately evaluating the memory kernel exists using
approximate simulations of
dynamics~\cite{mulvihillModifiedApproachSimulating2019,
mulvihillRoadMapVarious2021}, these are not controlled approximations. We, on
the other hand, use the framework of transfer tensor method
(TTM)~\cite{cerrilloNonMarkovianDynamicalMaps2014}, which decomposes the
dynamical maps, $\mathcal{E}(t)$, of the reduced system defined by $\tilde\rho(t) =
\mathcal{E}(t)\tilde\rho(0)$, in terms of transfer tensors $T_k$ defined as
\begin{align}
    T_k &= \mathcal{E}(k\Delta t) - \sum_{m\ge 1} T_m \,\mathcal{E}((k-m)\Delta t),
\end{align}
such that the time evolution of the system can be expressed as
\begin{align}
    \Tilde{\rho}(t_n) &= \sum_{k=1}^{L} T_k \, \Tilde{\rho}(t_{n-k})
\end{align}
where $L\Delta t=\tau_{\mathrm{mem}}$. These transfer tensors can now be related to the memory kernel as
\begin{align}
    T_k &= \mathcal{E}_0(\Delta t) \delta_{k,1}+\mathcal{K}_k\Delta t^2
\end{align}
where $\mathcal{E}_0 (t) = \exp\left(-iH_0 t / \hbar\right) \otimes \exp\left(iH_0 t / \hbar\right)$  is the dynamical map corresponding to the bare system, $H_0$.

To obtain the transfer tensors $T_k$, and consequently the memory kernels
$\mathcal{K}_k$, the dynamical maps, $\mathcal{E}(t)$, are simulated using path
integrals. These dynamical maps propagate the reduced density matrix of the
system in the presence of the environment. Suppose that at the initial time the system
is in a separable state with the environment in a thermal equilibrium, $\rho(0)
=
\tilde\rho(0)\otimes\exp(-\beta H_\text{env})/Z$. Then the dynamical map of the system in presence of the environment can be obtained using non-perturbative, numerically exact path integral methods:~\cite{makriTensorPropagatorIterativeI1995, makriTensorPropagatorIterativeII1995}
\begin{align}
    \tilde\rho(N\Delta t) &= \mathcal{E}(N\Delta t) \tilde\rho(0)\\
    \mel{s_N^\pm}{\mathcal{E}(N\Delta t)}{s_0^\pm} &= \sum_{s_1^\pm}\,...\sum_{s_{N-1}^\pm} \mel{s_N^\pm}{\mathcal{E}_0(\Delta t)}{s_{N-1}^\pm} \nonumber \\
    &\times \mel{s_{N-1}^\pm}{\mathcal{E}_0(\Delta t)}{s_{N-2}^\pm}\cdots\nonumber \\
    &\times \mel{s_1^\pm}{\mathcal{E}_0(\Delta t)}{s_0^\pm} F[\{s_i^\pm\}]
\end{align}
where $s_i^\pm$ is the state of the system at $i$th time point and
$F[\{s_i^\pm\}]$ is the Feynman-Vernon influence
functional~\cite{feynmanTheoryGeneralQuantum1963} along the path $\{s_i^\pm\}$.
The interactions between different time points caused by environmental
effects, which are encoded in the influence functional, are related to the
spectral density, Eq.~\ref{eq:spectral_density}, and the bath response
function~\cite{boseZerocostCorrectionsInfluence2022}.

With the memory kernels $\mathcal{K}_j$ in hand, we discretize Eq.~\ref{eq:piLD_diff} to finally get the reduced density matrix at the $n$th
time-step with the incorporation of loss mechanisms
\begin{align}
    \tilde\rho^{(L)}_n &= \mathcal{E}_0(\Delta t)\,\tilde\rho^{(L)}_{n-1}+\sum_{j=1}^{L}\mathcal{K}_j\tilde\rho^{(L)}_{n-j}\Delta t^2 \nonumber \\
    &+\sum_j \left(L_j\tilde\rho^{(L)}_{n-1}\,L_j^\dag - \frac{1}{2} \acomm{L_j^\dag L_j}{\tilde\rho^{(L)}_{n-1}}\right)\Delta t\label{eq:piLD}.
\end{align}

Now, we are in a position to simulate linear spectra of any open quantum system. A linear spectra is related to the Fourier transform of a system time correlation function of the form:
\begin{align}
    C_{A-B}(t) &= \Tr\left[A(t)B(0)\rho(0)\right],\\
    \text{where }\rho(0) &= \tilde\rho(0)\otimes\frac{e^{-\beta H_\text{env}}}{Z_\text{env}}
\end{align}
and $A$ and $B$ are system operators. One can define a time-evolved quantity
\begin{align}
    \rho_B(t) &= \Tr_\text{env}\left[\exp\left(-iHt/\hbar\right)B\rho(0)\exp\left(iHt/\hbar\right)\right]
\end{align}
which can be computed using PILD for an arbitrary environment. Now, the correlation function can be obtained as $C_{A-B}(t) = \Tr_\text{sys}\left[A\rho_B(t)\right]$. The subscripts, sys and env, denote partial trace over system and environment degrees of freedom respectively.

The two specific cases of linear spectra that are studied here are the absorption and the circular dichroism spectra. The absorption spectra is the real part of the Fourier transform of the transition dipole moment autocorrelation function ($A=B=\mu$)
\begin{align}
    \sigma_\text{abs}(\omega)  &\propto \Re\left[\int_0^t e^{i\omega t}C_{\mu-\mu}(t) \dd{t}\right] \label{eq:abs}
\end{align}
where $\mu=\sum_j\vb{\mu}_j$ is the total transition dipole moment constituted
by the individual molecular transition dipoles, $\vb{\mu}_j$. For a molecular
aggregate, the CD spectrum is defined in terms of the correlation function with
$A=m$ and $B=\mu$ as~\cite{cupelliniAbsorptionCircularDichroism2020}
\begin{align}
    \sigma_{\text{CD}}(\omega) &\propto \Im\left[\int_0^t e^{i\omega t}C_{\mu-m}(t) \dd{t}\right] \label{eq:CD}
\end{align}
where the magnetic moment of the aggregate, $m\propto\sum_j\vec{r}_j\cross\mu_j$
given that the $j$th molecule with a transition dipole of $\mu_j$ is located at
$\vec{r}_j$. The CD spectrum is instrumental in analysing orientational
chirality in molecular aggregates. Using the correlation function-based
definitions of the spectra allows for an exact simulation of the lineshape for a
particular environment at a particular temperature.

\begin{figure}
    \includegraphics{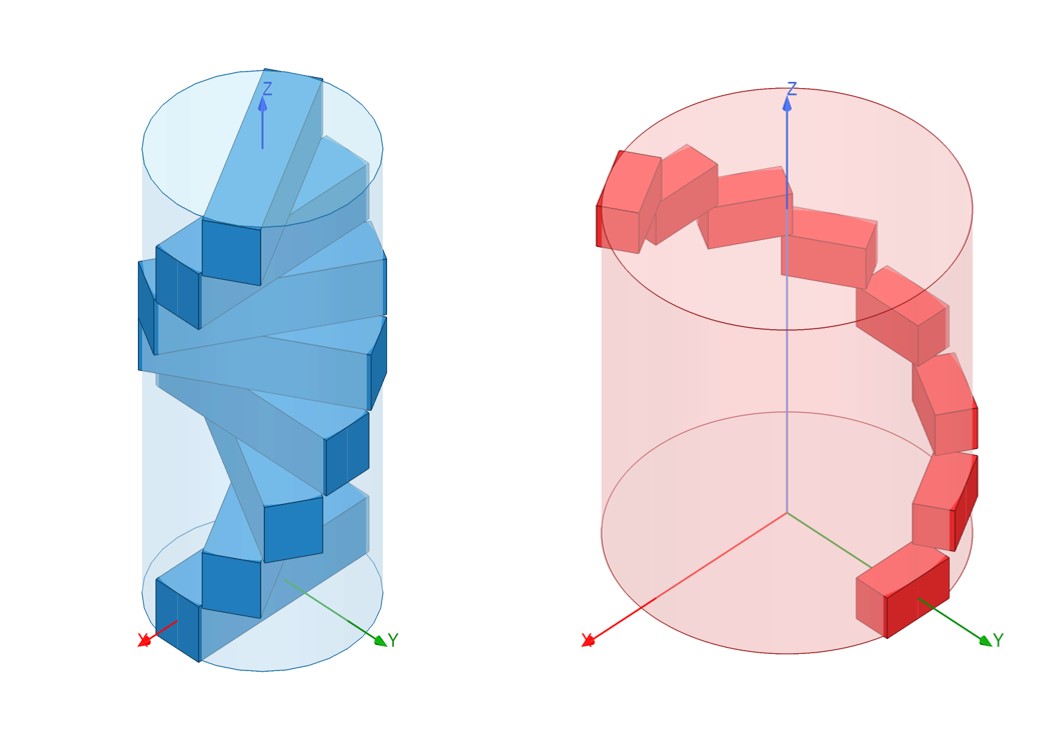}
    \caption{Geometries of molecular aggregates: Helix (right-handed) (left, blue) and Creeper (right-handed) (right, red)}\label{fig:geometry}
\end{figure}

\section{Numerical Results \& Analysis}\label{sec:results}
We will explore the spectra of two different classes of systems. The first being
that of an excitonic molecular aggregate, which is described by a
Frenkel-Holstein Hamiltonian given in Eq.~\ref{eq:Htot}. The corresponding system as defined by
Eq.~\ref{eq:H0_ex} has $N$ bacteriochlorophyll (BChl) molecules of equal monomer excitation energy
($\epsilon_j=\epsilon$ for all j) that interact in a nearest-neighbor fashion
($h_{j,k}=0$ for $\abs{j-k}>1$) in two different geometries, the right-handed
helix ($h_{j,j+1}=\SI{363}{\per\cm}$) and the right-handed creeper
($h_{j,j+1}=\SI{-363}{\per\cm}$). The consecutive individual molecular
transition dipoles of both arrangements are rotated counterclockwise by
$\frac{\pi}{6}$. For the helix, the molecular centers are all aligned along $z$-axis, whereas
the displacement vectors for consecutive molecules make an angle of
$\frac{\pi}{4}$ with the z-axis and the projections of their position vectors in
the $xy$-plane make an angle of  $\frac{\pi}{6}$ with each other for the creeper
geometry. The intermolecular distance for the creeper is 0.79 times that of the
helix. Both these arrangements are graphically depicted in Fig.~\ref{fig:geometry}.

\begin{figure}
    \includegraphics{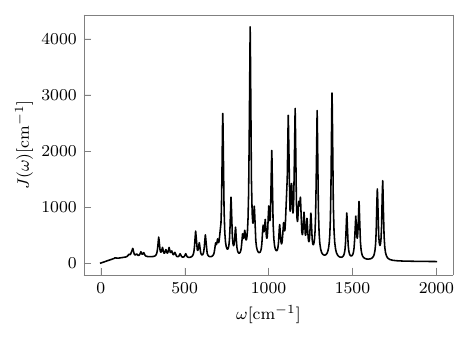}
    \caption{Spectral density corresponding to a bacteriochlorophyll molecule with reorganization energy of \SI{327}{\per\cm}.}\label{fig:bchl_spectral_density}
\end{figure}

The vibrations associated with the BChl monomers have been incorporated using the
Huang-Rhys factors corresponding to the interaction between 50 of the most
strongly coupled vibrations and the excited state of the molecules which have been
reported by R\"atsep \textit{et
al}~\cite{ratsepDemonstrationInterpretationSignificant2011}. These include only
relatively rigid molecular vibrations. To make the system more realistic, we
have augmented the spectral density using a Brownian bath described by a
Drude-Lorentz spectral density with a reorganization energy of \SI{109}{\per\cm}
and broadened the sharp Huang-Rhys factors by a Lorentzian of width
\SI{10.97}{\per\cm}. The resultant spectral density is shown in
Fig.~\ref{fig:bchl_spectral_density}. The temperature is set to $\SI{300}{\kelvin}$.

The second class of systems whose spectra we explore is exciton-polaritonic in
nature. The same BChl aggregates that were just described are put
inside a cavity. The system can now be written as
\begin{align}
    H_0 &= \sum_j\epsilon_j\dyad{j} + \sum_{j} h_{j,j+1}(\dyad{j}{j+1} + \dyad{j+1}{j})\nonumber\\
    &+ \hbar\omega_c\dyad{c} + \frac{1}{\sqrt{N}}\sum_j s_j(\dyad{c}{j} + \dyad{j}{c}),\label{eq:H0_pol}
\end{align}
where $\ket{c}$ is the state with all the molecules in the ground state and the cavity in the excited state, $\omega_c$ is the frequency of the cavity, and $s_j$ is the coupling of the $j$th monomer to the cavity. Assuming the electric field of the cavity mode to be aligned with transition dipole moment of the first monomer, we take $s_j=s\cos{\left[(j-1)\frac{\pi}{6}\right]}$ where $s=\SI{403.28}{\per\cm}$ and $\hbar\omega_c=\epsilon$. It is also important to note that the cavity is not coupled to a thermal environment.

Both the spectra are governed by correlation functions involving the total transition dipole moment of the aggregate. This is obtained as
\begin{align}
    \vb{\mu} &= \sum_j\va{\mu}_j\left(\dyad{j}{0} + \dyad{0}{j}\right).\label{eq:dipole}
\end{align}
Notice that the transition dipole moment operator does not involve the cavity.

We will explore the effect of loss from a particular monomer, $\ket{j}$, with a
decay time constant, $\tau$, on the absorption and circular dichroism spectrum
in various cases. This can be described by the Lindblad operator,
$L_j=\tau^{-1/2}\dyad{0}{j}$. Additionally, in the cases of
excitonic-polaritons, the cavities are generally leaky. This loss can be
captured by the corresponding Lindblad operator, $L_c=\tau_c^{-1/2}\dyad{0}{c}$.
The simulations are done using the path integral Lindblad
method~\cite{boseIncorporationEmpiricalGain2024} with the transfer tensors being
generated using the time-evolved matrix product operators (TEMPO)
algorithm~\cite{strathearnEfficientNonMarkovianQuantum2018} with multiple
baths~\cite{boseTensorNetworkRepresentation2021} as implemented in the
open-source QuantumDynamics.jl simulation
framework~\cite{boseQuantumDynamicsJlModular2023}. The inclusion of loss through
the PILD guarantees certain conserved quantities in the form of sum rules. The
area under the linear spectra, being related to the zero time value of
correlation functions, is invariant to the loss incorporated in the system.

\subsection{Spectroscopy of Excitonic Molecular Aggregates}
\begin{figure}
    \includegraphics{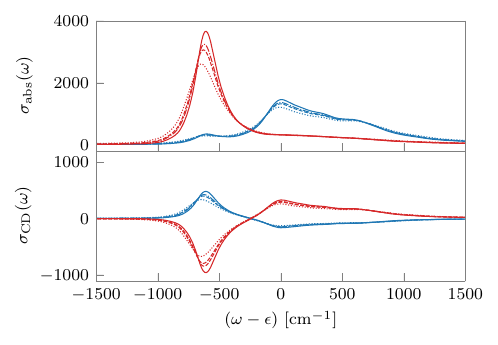}
    \caption{Absorption and CD spectra of BChl dimers in Helix (H-aggregate) (Blue) and Creeper (J-aggregate) (Red) geometries  for different decay constants $\tau=\SI{25}{\fs}$ (dotted), $\SI{50}{\fs}$ (dashed), $\SI{75}{\fs}$ (dash-dot), and the loss-less case (solid).}\label{fig:dimer}
\end{figure}

We start by exploring the spectra of the excitonic molecular aggregates under
losses with different decay times. Consider a BChl dimer in the creeper and the
helix formations. In Fig.~\ref{fig:dimer}, we show the absorption and CD spectra
for the system with decay times, $\tau=\SIlist{25; 50; 75}{\fs}$ along with that
of the loss-less dimer (in solid curve). The profile of the absorption peak of
the helix is significantly more structured and broader than that of the creeper.
It is usual for H-aggregates to have more broadened spectra compared to the
J-aggregates when coupled with structured vibrational bath.~\cite{spanoHJAggregateReview2018} The CD spectrum
shows a positive first and a negative second Cotton feature for the helix
geometry which is opposite to that of the creeper geometry despite both
geometries being right-handed. It is notable that, while the two expected peaks
corresponding to the two bright excitonic states for a noncoplanar dimer are
not easily discernible from the absorption spectrum (especially for the
creeper), it is possible to clearly locate them from the CD spectrum.

In Fig.~\ref{fig:dimer}, it is clearly seen that though both the helix and the
creeper geometries have the same absolute chirality, the CD spectra have
different Cotton effects. This is in apparent contradiction of the often-used
exciton-chirality (EC) rule, which uniquely maps the sign of the lowest-energy
Cotton couplet to the absolute chirality of the
aggregate~\cite{haradaExcitonChirality1975, haradaECDPrinciples2012,
harada1968optical}. However, Swathi \textit{et
al.}~\cite{sissaCaveatinEXcitonChiralityRule2020} have shown that, consistent
with our results here, the EC mapping is reversed between the H- and J- dimers. 

On applying the different losses, the spectral feature of the dimer broaden and
become less intense. Notice that the relative decrease in the intensity of the
primary absorption peak with respect to the loss-less case is more for the
creeper than for the helix. This broadening is also accompanied by an increasing
red shift of the spectra. (This is also seen in the absence of the dissipative
environment as demonstrated in Appendix~\ref{app:lindblad}.) The red shift of the
$\omega<\epsilon$ peak seems to be higher than that of the $\omega\ge\epsilon$
peak. In addition, the relative magnitude of attenuation of the peak height of
the helix is much smaller than that of the creeper. These observations also
extend to the CD spectrum where the shift and attenuation due to loss mechanisms
are more prominent for the sharper peaks (or troughs). 

\begin{figure}
    \centering
    \subfloat[Helix geometry]{\includegraphics{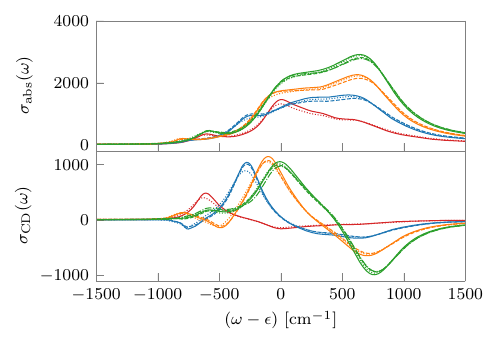}}
    
    \subfloat[Creeper geometry]{\includegraphics{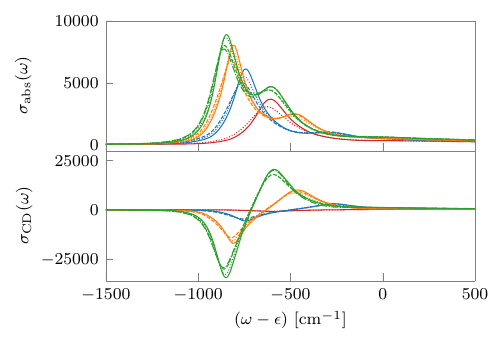}}
    \caption{Absorption and CD spectra of BChl Dimer (Red), Trimer (Blue), Tetramer (Orange) and Pentamer (Green) for decay constant $\tau=\SI{50}{fs}$ at monomer 1 (dotted), monomer 2 (dashed), monomer 3 (dash-dot) and the loss-less cases (solid).}
    \label{fig:excitonic-aggregates}
\end{figure}

From the dimer as a representative case, we proceed to investigate the effect of
exciton system size vis-à-vis decay time $\tau$ in
Fig.~\ref{fig:excitonic-aggregates}~(a) and~(b). We look at the dimer, trimer,
tetramer, and pentamer with the same nearest-neighbor intermolecular couplings.
Losses with a decay time of $\tau=\SI{50}{fs}$ are applied to each of the unique
sites in the aggregate. For the trimer and tetramer, monomers 1 and 2 constitute
the two distinct sites of decay while for the pentamer, monomer 3 is an
additional unique site. While we observe the expected red shift in the
absorption spectra for the J-creepers (Fig.~\ref{fig:excitonic-aggregates}~(b))
for all system sizes, there is an increasing blue shift for the trimer H-helices
and beyond (Fig.~\ref{fig:excitonic-aggregates}~(a)). The impact of a loss on a
single site is seen to decrease with increasing system size. Moreover, decay at
the central monomer is more effective in attenuating the intensity of spectrum
than the decay at other site owing to symmetries of the system.

The CD spectra shown in Fig.~\ref{fig:excitonic-aggregates}~(a) and~(b) enables
identification of various peaks obscured in the absorption spectrum. For the
J-creepers, the sign of the first peak is consistently seen to be negative at
all sizes of the aggregates. However, it is interesting that all the H-helices
do not show the same sign of the first peak of the CD spectra. For the case of
the helix trimer, one sees the initial peak become negative instead of the
positive peak seen in the other helices. This further shows that the EC mapping
between the CD spectra and the corresponding geometry is even more tenuous.
Earlier we had discussed the inconsistency between a helix and creeper dimer.
Now we note the inconsistency between helices of different sizes. However, this pattern
can be explained for the bare system through a calculation of the
rotational strengths of each of the excitonic states as outlined in
Appendix~\ref{app:rot}. Also, as seen in the case of the isolated pentamer, for
both the creeper and the helix, there are only three states that are CD active.

\subsection{Spectroscopy of Excitonic-Polaritonic Systems}
\begin{figure}
    \includegraphics{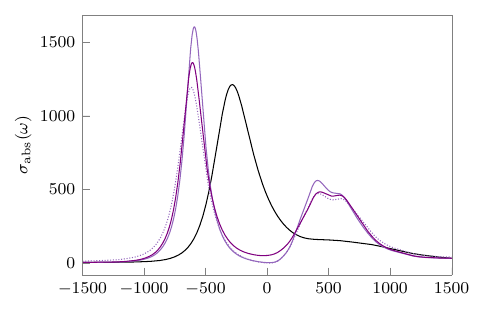}
    \caption{Absorption spectrum of BChl Monomer with (Purple) and without (Black) an Optical Cavity for decay constant $\tau=\SI{50}{\fs}$ at monomer (dotted), cavity (solid; darker shade) and the loss-less case (solid; lighter shade).}\label{fig:monomercav}
\end{figure}

As a second class of systems of interest, we consider excitonic systems coupled
with a plane-polarized cavity mode. In Fig.~\ref{fig:monomercav}, we see how
upon coupling to the cavity mode, a monomer absorption peak splits into two
polaritonic peaks corresponding to the lower energy ``Lower Polariton''
and higher energy ``Upper Polariton'' which are the two
eigenstates of Eq.~\ref{eq:H0_pol}. In absence of any loss, the upper polariton shows a broader line-shape, which correlates with a lower lifetime in comparison to the lower polariton. We notice that similar to the excitonic
systems, the decrease in intensity and broadening of peaks on the inclusion of
loss mechanisms are not equal. It is more for the lower polariton than the upper
polariton. This difference, while relatively minor when the cavity is lossy,
becomes clearly noticeable when the loss is on the monomer. This
is entirely due to interaction of the environment with the different polaritons
(Appendix~\ref{app:lindblad}). However, unlike the excitons, where loss
mechanisms always led to a red shift in the absorption spectrum, the lower
polariton $\ket{L}$ experiences a red shift while the upper polariton, $\ket{U}$
a blue shift. This effective expansion of the polaritonic spectrum is opposite
to the compression observed in the absence of the dissipative environment as
demonstrated in Appendix~\ref{app:lindblad}. Interestingly, a similar effect has
been noticed using non-Hermitian systems to model the dissipation and the
so-called $P(E)$ theory~\cite{kansanenPolaritonResponsePresence2021}. Here we
can quantify the non-perturbative effects of the thermal environment in a
numerically exact manner. Thus, while the lower polariton is energetically
stabilized by the external loss, its lifetime decreases more prominently than
the upper polariton.

\begin{figure}
    \centering
    \subfloat[Helix geometry]{\includegraphics{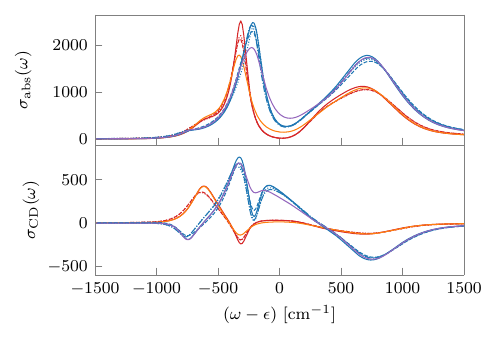}}
    
    \subfloat[Creeper geometry]{\includegraphics{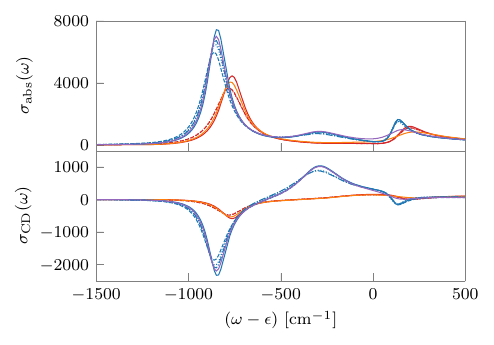}}
    \caption{Absorption and CD spectra of BChl Dimer (Red) and Trimer (Blue) in an optical cavity for decay constant $\tau=\SI{50}{fs}$ at monomer 1 (dotted), monomer 2 (dashed), monomer 3 (dash-dot), cavity (dimer: orange, trimer: purple) and the loss-less cases (solid).}
    \label{fig:excav-aggregates}
\end{figure}

Finally, in Fig.~\ref{fig:excav-aggregates}~(a) and~(b), we demonstrate the
absorption and CD spectra for BChl dimer and trimer coupled to a cavity mode.
While a polaritonic system described by the Holstein-Tavis-Cummings model has
only two bright states~\cite{herreraCavityControlledChemistry2016}, here we have
more. This apparent discrepancy stems from the Holstein-Tavis-Cummings model
ignoring the intermolecular coupling. We notice in both geometries that the
effect of a loss at the cavity is qualitatively different from that at the
monomers. For the helix geometry, the lowest and highest energy CD peaks seem to
remain relatively unaffected by a cavity loss. The intermediate peaks are all
highly damped. The effect seems to be reversed for the creeper geometry
(Fig.~\ref{fig:excav-aggregates}~(b)). Thus, there is a more marked decrease in
lifetime of particular polaritons due to the loss mechanisms. Additionally, it
is interesting that the first CD couplet resembles the excitonic system thereby
aiding in identification of the geometry of the aggregate. The CD spectra
actually provides a way to identify the dark states as the peaks (or troughs)
which are least affected by loss at cavity because of the small contribution of
cavity mode to these states.

The interpretation of the CD spectra, the changes that happen in them due to
loss, and correlating it with structure of the polaritonic system is a highly
nuanced endeavor. This will be explored in the future in greater detail.

\section{Conclusion}\label{sec:conclusions}
In this paper, we present a numerically rigorous study of the impact of
empirically described loss mechanisms on two different linear spectra,
absorption spectra and circular dichroism spectra, of condensed phase systems
using the recently developed path integral Lindblad dynamics method. Empirical
loss mechanisms are often incorporated using non-Hermitian descriptions which
make the time-evolution non-unitary. This has spurious effects on the spectra.
PILD ensures that the dynamical maps are unitary and consequently avoids such
spurious effects. An additional benefit of PILD is that one can get the effect
of these loss mechanisms at no extra cost, while maintaining the accuracy of the
method and the rigor of accounting for the effects of the thermal dissipative
environment. One of the consequences of the accuracy of these simulations is
that the area under the spectra are invariant under changing the time-scale of
the losses.

Intuitively, one might think that the effect of loss is the broadening of the
spectrum. This is indeed the primary effect. However, the details are much more
complex. Different peaks get broadened in different ways. The empirical loss
also ``interacts'' with the dissipative environment to have more complex and
apparently unpredictable amounts of loss. Additionally, this broadening can also
be accompanied by a shift of the peak. We demonstrate these complexities through
various examples.

We consider the absorption and the CD spectra of chiral excitonic aggregates and
chiral polaritonic aggregates formed by these systems inside a nonchiral,
plane-polarized cavity. The numerically exact spectra of the exictonic
aggregates are already quite complicated. We show that different peaks get
attenuated at different rates. The magnitude of the lossy site to the particular
exciton, while an important consideration, is not the only factor which
determines the amount by which the peak would get attenuated. There are complex
interactions mediated by the environment which couples different excitons
together. Additionally, we see that the helix aggregate with primarily H-type
interactions have far broader peaks than the creeper aggregate with J-like
interactions. This greater structure has typically been attributed to
exciton-phononic bands~\cite{spanoHJAggregateReview2018} for simple cases.

The presence of the cavity adds to the complexity of the problem. The excitonic
absorption peaks get split into multiple polaritonic peaks. We report probably the first numerically exact path integral simulations of CD spectra of chiral
aggregates in cavity. The chirality of the different polaritonic peaks interacts
in unexpected manners. For a monomer or a dimer for instance, the upper
polariton gets attenuated significantly less than the lower polariton.

All these complexities and subtleties necessitate rigorous simulation
frameworks for predicting the spectra in these open quantum systems. We provide
the first exploration of such effects by using the rigorous and fully quantum
mechanical PILD framework based on TEMPO QuAPI calculations. The ability to
understand and predict the spectra provides a powerful handle on the properties of these
complex aggregates. This would be a topic of exploration and study in the
future. Further these numerically exact methods will be used to elaborate the
routes of transport and their relative efficiencies in a future work.

\appendix
\section{Spectra of lossy systems without vibrational decohering environments}\label{app:lindblad}
To understand the changes better, we demonstrate the impact of the loss
analytically on the spectra of a linear (co-planar) excitonic dimer and a
monomer interacting with a cavity mode without any thermal environment. This can
be done analytically and enables separation of impact caused solely by the loss
versus that caused through the mediation of the environment.

Consider first the excitonic dimer without the dissipative environment as
defined in Eq.~\ref{eq:H0_ex} with equal monomer excitation energy $\epsilon$
and intermonomer coupling as $h_{12}(=h_{21})$. As in the cases described
earlier, there is a loss with a time-scale of $\tau$ on the second monomer given
by $L_2=\tau^{-1/2}\dyad{0}{2}$. For simplicity of analysis, we assume that the
decay timescale is relatively weak, $\tau>\frac{\hbar}{4h_{12}}$. This would
typically be the case because the electronic couplings between the monomers
would be stronger than the time-scales of a loss process. The transition dipole
moment operator is defined in Eq.~\ref{eq:dipole} with equal magnitudes $\mu$
for each monomer. 
%One can analytically calculate the correlation function which comes out to be a function of $\tau$:
%\begin{align}
%    \scriptstyle C_{\mu-\mu}^{\boldsymbol{(2)}}(t) &= \scriptstyle \hspace{0.3em}\mu^2 e^{-i\epsilon t/\hbar}\left[(e^{i\alpha t}+e^{i\beta t})+\frac{2h_{12}}{\hbar(\beta-\alpha)}(e^{i\alpha t}-e^{i\beta t})\right]
%\end{align}
%where for $a\in\mathbb{R}$:\vspace{-1em}
%\begin{align}
%    \scriptstyle \alpha &= \scriptstyle \hspace{0.3em}\frac{i}{4\tau}-\sqrt{\frac{h_{12}^2}{\hbar^2}-\frac{1}{16\tau^2}} \hspace{2em}(= \scriptstyle\,ib-a) \label{eq:alpha}\\
%    \scriptstyle \beta &= \scriptstyle \hspace{0.3em}\frac{i}{4\tau}+\sqrt{\frac{h_{12}^2}{\hbar^2}-\frac{1}{16\tau^2}} \hspace{2em}(= \scriptstyle\,ib+a)\label{eq:beta}
%\end{align}
%Following the relation~\ref{eq:abs}, we get:
The absorption spectrum can now be calculated using the relation~\ref{eq:abs} to get:
\begin{align}
%    \scriptstyle \sigma_{abs}^{\boldsymbol{(2)}}(\omega) &\propto \scriptstyle \hspace{0.3em}\Re\left[\frac{i\mu^2}{\alpha-\beta}\left(\frac{\alpha-\beta-2h_{12}/\hbar}{\omega+\alpha-\epsilon/\hbar} +\frac{\alpha-\beta+2h_{12}/\hbar}{\omega+\beta-\epsilon/\hbar}\right)\right] \nonumber \\
    \sigma_\text{abs}^{\boldsymbol{(2)}}(\omega)
    &\propto \mu^2\frac{b}{a}\left[\frac{a+h_{12}/\hbar}{\left(\omega-a-\epsilon/\hbar\right)^2+b^2}+\frac{a-h_{12}/\hbar}{\left(\omega+a-\epsilon/\hbar\right)^2+b^2}\right]\label{eq:dimer_analytic}
\end{align}
where for $a=\sqrt{\frac{h_{12}^2}{\hbar^2}-\frac{1}{16\tau^2}}$ and
$b=\frac{1}{4\tau}$. Notice that this is a sum of two Lorentzians centered at
$\omega_0 = a+\epsilon/\hbar$ and $\omega_1 = -a+\epsilon/\hbar$ respectively.
If $\tau\gg\frac{\hbar}{4h_{12}}$, then the intensity of the peak at $\omega_1$
would be very small and can be neglected. Consequently on decreasing $\tau$
(i.e., a stronger decay), the dominant peak at $\omega_0$ shows a red-shift
owing to the decrease in the value of $a$.

Similarly, for a monomer-cavity system defined with Eq.~\ref{eq:H0_pol} with the
cavity resonant with monomer excitation energy, $\hbar\omega_c=\epsilon$, cavity
coupling $s$ and transition dipole moment as defined in Eq.~\ref{eq:dipole} with
magnitude $\mu$, we consider the two cases with the loss on the cavity
$L_c=\tau^{-1/2}\dyad{0}{c}$ and a loss on the monomer
$L_1=\tau^{-1/2}\dyad{0}{1}$. The corresponding absorption spectra for the loss
on the cavity and the monomer are: %The correlation functions and corresponding absorption spectrum for loss on cavity are as follows:
\begin{widetext}
\begin{align}
%    \scriptstyle C_{\mu-\mu}^{\boldsymbol{(c)}}(t) &= \scriptstyle \hspace{0.3em}\mu^2 e^{-i\epsilon t/\hbar}\left[i\frac{\left(e^{i\alpha' t}-e^{i\beta' t}\right)}{4\tau(\beta'-\alpha')}+\frac{1}{2}\left(e^{i\alpha' t}+e^{i\beta' t}\right)\right] \\
%    \scriptstyle \sigma_{abs}^{\boldsymbol{(c)}}(\omega) &\propto \scriptstyle \hspace{0.3em}\Re\left[\frac{\mu^2}{4\tau(\alpha'-\beta')}\left(\frac{2i\tau(\alpha'-\beta')+1}{\omega+\alpha'-\epsilon/\hbar} +\frac{2i\tau(\alpha'-\beta')-1}{\omega+\beta'-\epsilon/\hbar}\right)\right]\nonumber \\
    \sigma_\text{abs}^{\boldsymbol{(c)}}(\omega)
    &\propto \mu^2\frac{b}{2a'}\left[\frac{2a'-\omega+\epsilon/\hbar}{\left(\omega-a'-\epsilon/\hbar\right)^2+b^2}+\frac{2a'+\omega-\epsilon/\hbar}{\left(\omega+a'-\epsilon/\hbar\right)^2+b^2}\right]\label{eq:cav_analytic} \\
%\end{align}
%Similarly, the correlation functions and corresponding absorption spectrum for loss on monomer are as follows:
%\begin{align}
%    \scriptstyle C_{\mu-\mu}^{\boldsymbol{(1)}}(t) &= \scriptstyle \hspace{0.3em}\mu^2 e^{-i\epsilon t/\hbar}\left[-i\frac{\left(e^{i\alpha' t}-e^{i\beta' t}\right)}{4\tau(\beta'-\alpha')}+\frac{1}{2}\left(e^{i\alpha' t}+e^{i\beta' t}\right)\right] \\
%    \scriptstyle \sigma_{abs}^{\boldsymbol{(1)}}(\omega) &\propto \scriptstyle \hspace{0.3em}\Re\left[\frac{\mu^2}{4\tau(\alpha'-\beta')}\left(\frac{2i\tau(\alpha'-\beta')-1}{\omega+\alpha'-\epsilon/\hbar} +\frac{2i\tau(\alpha'-\beta')+1}{\omega+\beta'-\epsilon/\hbar}\right)\right]\nonumber \\
    \sigma_\text{abs}^{\boldsymbol{(1)}}(\omega)
    &\propto \mu^2\frac{b}{2a'}\left[\frac{\omega-\epsilon/\hbar}{\left(\omega-a'-\epsilon/\hbar\right)^2+b^2}-\frac{\omega-\epsilon/\hbar}{\left(\omega+a'-\epsilon/\hbar\right)^2+b^2}\right]\label{eq:monomer_analytic}
    %&=\frac{2\mu^2b\left(\omega - \epsilon/\hbar\right)}{\left(\omega - \epsilon/\hbar\right)^2 - a^2 + b^2}\\
\end{align}
\end{widetext}
Here, $a'=\sqrt{\frac{s^2}{\hbar^2}-\frac{1}{16\tau^2}}$, which is the same as $a$ but with $h_{12}$ replaced by $s$ while the symbol in superscript parenthesis denotes the site of loss.

We can now see from the relation~\ref{eq:cav_analytic} for the loss on cavity that instead of shifting in the same direction, the lower energy polariton shifts blue, whereas the higher energy polariton shifts red. This is exactly opposite to what we saw in presence of the dissipative media (Fig.~\ref{fig:monomercav}). However with relation~\ref{eq:monomer_analytic} for the loss on monomer, there is no conspicuous shift in the spectrum. It is also to be noted that the lineshape for the two polaritonic peaks are symmetrical despite the inclusion of loss at any site.

\section{Rotatory strengths for peaks of excitonic aggregates}\label{app:rot}
Here we discuss the rotatory matrix and the strengths for bare excitonic and polaritonic aggregates in absence of thermal dissipative environments. Consider a general aggregate, either excitonic or polaritonic, described by a Hamiltonian, $\hat{H}_0$, with $N$ molecules located at $\va{r}_j$ with transition dipoles $\va{\mu}_j$. The cross correlation function between the transition dipole moment and the magnetic moment can be refactored as follows:
\begin{align}
    \sigma_\text{CD}(\omega) &= \Re\int_0^\infty \sum_{ij} R_{ji} I_{ij}(t) \exp(i\omega t)
\end{align}
where:
\begin{align}
     R_{ji} &= -\sqrt{\epsilon_i\epsilon_j} \left(\va{r}_j - \va{r}_i\right) \cdot \left(\va{\mu}_i \cross \va{\mu}_j\right)\\
    I_{ij}(t) &= e^{iE_gt/\hbar}\Tr_\text{env}\left(\mel{j}{e^{-iHt/\hbar}}{i}\right)
\end{align}
The rotatory strength matrix is $R$ and $I(t)$ is the lineshape
function. Now, this is defined in the molecular basis. We want the strengths
corresponding to the different eigenstates of the system Hamiltonian
$\hat{H}_0$, $\ket{e_j}$. These states, $\ket{e_j}$, correspond to excitons in
the molecular aggregates and polaritons in presence of cavities. It is possible to transform the spectrum from a molecular basis to the eigenstate basis:
\begin{align}
    \sigma_\text{CD}(\omega) &= \Re\int_0^\infty \Tr\left(R I(t)\right)\exp(i\omega t)\\
    &= \Re\int_0^\infty \sum_{ij}\mel{e_i}{R}{e_j}\mel{e_j}{I(t)}{e_i}\exp(i\omega t)
\end{align}
For an isolated system in absence of dissipative environment, the transformation is
simple. The lineshape function matrix is diagonal in the $\ket{e_j}$ basis. It
is the presence of the dissipative environment that couples the different
eigenstates. Consequently, the peak heights for an isolated system are
proportional to the matrix element of the rotatory strength matrix corresponding
to the particular eigenstate, $\mel{e_j}{R}{e_j}$.
\begin{align}
    \sigma_\text{CD}(\omega) &= \sum_i \mel{e_i}{R}{e_i}\delta(\omega - E_i - E_g).
\end{align}

\begin{table*}
    \centering
    \begin{tabular}{cc||cc||cc||cc}
        \multicolumn{2}{c}{Dimer} & \multicolumn{2}{c}{Trimer} & \multicolumn{2}{c}{Tetramer} & \multicolumn{2}{c}{Pentamer}\\
        $\omega' (\unit{\per\cm})$ & $\sigma_\text{CD}(\omega)$ & $\omega' (\unit{\per\cm})$ & $\sigma_\text{CD}(\omega)$ & $\omega' (\unit{\per\cm})$ & $\sigma_\text{CD}(\omega)$ & $\omega' (\unit{\per\cm})$ & $\sigma_\text{CD}(\omega)$\\\hline\hline
        $-363.0$ & $-0.5$ & $-513.36$ & $-1.57$ & $-587.35$ & $-3.19$ & $-628.73$ & $-5.20$\\
        $+363.0$ & $+0.5$ & $0.0$ & $+1.73$ & $-224.35$ & $+3.41$ & $-363.0$ & $+5.10$\\
         &  & $+513.36$ & $-0.16$ & $+224.35$ & $-0.31$ & $0.0$ & $0.0$\\
         &  &  &  & $+587.35$ & $+0.09$ & $+363.0$ & $+0.10$\\
         &  &  &  &           &         & $+628.73$ & $0.0$
    \end{tabular}
    \caption{CD peak heights for isolated creeper aggregate as a function of aggregate size. ($\omega'=\omega-\epsilon$)}
    \label{tab:creeper}
\end{table*}
\begin{table*}
    \centering
    \begin{tabular}{cc||cc||cc||cc}
        \multicolumn{2}{c}{Dimer} & \multicolumn{2}{c}{Trimer} & \multicolumn{2}{c}{Tetramer} & \multicolumn{2}{c}{Pentamer}\\
        $\omega' (\unit{\per\cm})$ & $\sigma_\text{CD}(\omega)$ & $\omega' (\unit{\per\cm})$ & $\sigma_\text{CD}(\omega)$ & $\omega' (\unit{\per\cm})$ & $\sigma_\text{CD}(\omega)$ & $\omega' (\unit{\per\cm})$ & $\sigma_\text{CD}(\omega)$\\\hline\hline
        $-363.0$ & $+0.5$ & $-513.36$ & $-0.16$ & $-587.35$ & $+0.09$ & $-628.73$ & $0.0$\\
        $+363.0$ & $-0.5$ & $0.0$ & $+1.73$ & $-224.35$ & $-0.31$ & $-363.0$ & $+0.10$\\
         &  & $+513.36$ & $-1.57$ & $+224.35$ & $+3.41$ & $0.0$ & $0.0$\\
         &  &  &  & $+587.35$ & $-3.19$ & $+363.0$ & $+5.10$\\
         &  &  &  &           &         & $+628.73$ & $-5.20$
    \end{tabular}
    \caption{CD peak heights for isolated helix aggregate as a function of aggregate size. ($\omega'=\omega-\epsilon$)}
    \label{tab:helix}
\end{table*}

In Tables~\ref{tab:creeper} and~\ref{tab:helix}, we list the
locations and the heights of the CD peaks for different sized nearest neighbor
creeper and helix aggregates with $\abs{h_{12}}=\SI{363.0}{\per\cm}$. Notice
that in absence of the thermal environment, the peak heights are exactly
reversed between the helix and the creeper. The locations are the same. With the
size of the aggregate, the height of the peak increases in case of the creeper.
However, there is no apparent pattern for the helix. With regards to the sign of
the first peak, notice that just as seen in presence of the environment,
irrespective of the aggregate size, the creeper has a negative peak first. We
see that the sign of the first peak of the helix keeps alternating between a
positive and a negative sign. This apparently simple pattern for helices is
further complicated by the fact that some times, like in the case of a pentamer,
the first peak has zero intensity. In that case, effectively, the first peak
that is symmetry allowed would be positive again. These patterns are consistent
with the spectra calculated in the presence of the solvent.

\bibliography{library}
\end{document}